# Enhanced Hot-Carrier Luminescence in Multilayer Reduced Graphene Oxide Nanospheres

Qi Chen[1], Chunfeng Zhang[1,*], Fei Xue[1], Yong Zhou[1], Wei Li[1], Ye Wang[1], Wenguang Tu[1], Zhigang Zou[1], Xiaoyong Wang[1], & Min Xiao[1,2,*]

[1]National Laboratory of Solid State Microstructures and Department of Physics, Nanjing University, Nanjing 210093, China, [2]Department of Physics, University of Arkansas, Fayetteville, Arkansas 72701, USA

Correspondence and requests for materials should be addressed to
C.Z. ( cfzhang@nju.edu.cn) or M.X.( mxiao@uark.edu)

**We report a method to promote photoluminescence emission in graphene materials by enhancing carrier scattering instead of directly modifying band structure in multilayer reduced graphene oxide (rGO) nanospheres. We intentionally curl graphene layers to form nanospheres by reducing graphene oxide with spherical polymer templates to manipulate the carrier scattering. These nanospheres produce hot-carrier luminescence with more than ten-fold improvement of emission efficiency as compared to planar nanosheets. With increasing excitation power, hot-carrier luminescence from nanospheres exhibits abnormal spectral redshift with dynamic feature associated to the strengthened electron-phonon coupling. These experimental results can be well understood by considering the screened Coulomb interactions. With increasing carrier density, the reduced screening effect promotes carrier scattering which enhances hot-carrier emission from such multilayer rGO nanospheres. This carrier-scattering scenario is further confirmed by pump-probe measurements.**

The potential of using graphene in photonics and optoelectronics has been realized by recent progresses in developing various graphene-based devices[1, 2] including optical modulators[3], light-emitting diodes[4, 5], photocatalysis[6-8], ultrafast lasers[9-11], solar cells[12, 13], and photodetectors[14, 15]. Nevertheless, the gapless feature of its band structure makes graphene an inefficient light emitter (with quantum yield < $10^{-12}$)[16]. The photo-excited hot carriers cool down to the lattice temperature before the relatively slow radiative recombination takes place due to the electron-electron (*e-e*) and electron-phonon (*e-ph*) interactions[16-21]. In order to promote light emission efficiency, various luminescent graphene derivatives have been synthesized by using advanced material processes in the past few years[22-32]. Some of these graphene forms, especially the graphene oxide (GO) and reduced GO (rGO)[24-32], can significantly enhance photoluminescence (PL) emission. Nevertheless, their electronic structures are normally modified to be different from pristine graphene, hindering certain potential applications that rely on the high mobility of graphene.

Recently, a pulse-pumping approach has been introduced to improve light emission with the aid of hot carriers in graphene[16,33-36]. Benefiting from carrier scatterings, a portion of hot carriers excited by ultra-short laser pulses recombine radiatively prior to the energy loss to phonon modes, so that the PL emission can be substantially enhanced[16, 33]. With this concept, Li *et al.* reported interesting phenomena of population inversion and optical gain in graphene, opening opportunity to explore new type of graphene lasers[35]. Of equal prominence is the earlier observation of nonlinear hot-carrier emission[16, 33]. Broadband emission from graphene layers was recorded in the up-converted frequency domain with over three orders of magnitude increase in emission efficiency as compared to the case with a continuous-wave excitation, which is promising for optical labeling and imaging. Hot-carrier emission is tightly associated with carrier dynamics in graphene which is a subject under extensive study in recent years. In theory, the dynamic behaviors of hot carriers, like carrier scattering, carrier-phonon interaction or even three-body interactions (supercollision cooling), can be tailored by multiple variables such as carrier density, defect levels, ripples, and so on[37-43]. Some of these predications have already been realized in recent experiments[44-46].

In this work, we attempt to promote nonlinear hot-carrier emission by manipulating carrier scattering in graphene materials. For this purpose, we introduce curvature to graphene layers and study the properties of hot-carrier luminescence in spherical nanostructures with femtosecond (fs) pulse excitation. We find that the rGO nanospheres, prepared by reducing GO with spherical templates, can produce hot-carrier luminescence with more than ten-fold improvement in emission efficiency as compared to the planar nanosheets. With increasing excitation power, a redshift of PL emission from nanospheres is observed, which is in contrary to the blueshift observed in planar layers[16, 33]. Temporal evolution of PL dynamics, as characterized by two-pulse correlation, shows biexponential decays with a fast component (< 0.2 ps) and a weak slow component (~0.8 ps). The amplitude of the slow component associated to the *e-ph* coupling increases dramatically with the rise of incident power. By analyzing the data with a hot-carrier emission model[33], we find that the screening effect is significantly reduced with increasing carrier density in rGO nanospheres. The enhanced scattering due to the curvature might be a plausible origin for the luminescence enhancement in such rGO nanospheres. Such a carrier-scattering scenario is further confirmed by the pump-probe study on the carrier dynamics in rGO nanospheres.

**Results**

**The morphologies of rGO nanospheres.** Hot-carrier emission in graphene materials is associated to the *e-e* and *e-ph* interactions[16, 33]. To promote hot-carrier emission, one can intentionally curl the graphene sheets to manipulate such many-particle interactions as theoretically discussed in previous literatures[37, 38]. We employ the technique of reducing GO with spherical polymer templates to synthesize the graphene hollow nanospheres as illustrated in Figure 1a[6]. In brief, we prepare precursors by modifying poly(methly methacrylate) (PMMA) spheres with a protonic polyethylenimine (PEI) aqueous solution and a negatively charged GO suspension. The precursors of core/shell composites, formed by electrostatic interaction, are then heated by microwave to fully reduce GO into graphene and to decompose the PMMA templates. The hollow rGO nanospheres are produced after removing the PMMA residue with tetrahydrofuran. In our previous study[6], we have shown that this procedure can produce spherical graphene nanostructures with crystalline quality comparable to the planar graphene samples prepared by reducing GO. As shown in Figure 1b, the samples are in the form of hollow spheres with an average diameter of ~ 300 nm as characterized by scanning and transmission electron microscopy. The thickness of the samples, which is controlled by the amount of GO in the precursor, is ~ 5 layers in this study. To find the unique properties associated with such spherical morphology, we conduct the study on the samples of rGO nanospheres and planar nanosheets comparatively. Raman spectrum of the nanospheres (Figure 1c) shows the signature modes including the G mode (~ 1585 cm$^{-1}$) and defect D mode (~ 1360 cm$^{-1}$). This Raman spectrum is comparable to that of the reference sample and rGO samples employed in the literature[47], indicating that the nanospheres are formed by the bonding between *sp$^2$* carbon sites. We notice that the Raman shift of 2D-mode in rGO nanospheres exhibits slight difference at different sample spots (inset, Figure 1c). It has been established that the 2D-mode Raman peak recorded from multilayer graphene samples is sensitive to the stacking order[48]. The observation here can be regarded as a signature of the random stacking order with certain spatial heterogeneity in rGO nanospheres, which is reasonable since the samples were prepared initially from monolayer GO nanosheets.

**Hot-carrier luminescence.** With fs pulse excitation at 800 nm, the graphene nanospheres produce efficient emission in the visible range (400 - 750 nm) (Figure 2a). The integrated emission intensity is superlinearly dependent on excitation fluence (*P*) as $P^{2.7}$ (Figure 2b). Under short-wavelength excitation, PL from GO or rGO samples is frequently observed with visible peak (Figure 2c), which is attributed to the localized states[27, 31]. Intuitively, the up-converted emission observed here may come from similar localized states populated by multi-photon absorption[30]. However, the emission spectra from nanospheres with no visible peaks are distinct from the PL spectra recorded in either GO or rGO samples (Figure 2c). This explicit difference suggests that other origin for the up-converted emission should be involved. We notice that the spectral dispersion is analogous to that of hot-carrier luminescence from exfoliated graphene layers[16, 33]. Also, the dependence of emission intensity on the incident power is very similar to the hot-carrier luminescence. As schematically shown in Figure 3a, hot-carrier luminescence in graphene is closely associated with carrier scattering. The scatterings between photo-induced carriers drive the electrons to higher energy states. Radiative recombination from these states induces hot-carrier luminescence in the up-converted frequency domain.

Hot-carrier scattering in graphene materials has been frequently considered as an example of the screened Coulomb interaction[37, 38, 41-43]. Within this framework, spectral dispersion of the emission can be modeled theoretically. Liu *et al.* derived a model under the Weisskopf-Wigner approximation, which predicts the spectral profile in the form of[33]

$$I(\omega) \propto \frac{\omega[\omega_0^2 - (\omega - \omega_0)^2]}{[\omega - \omega_0 + v_F Q_{TF}]^4}, \quad (1)$$

where $\hbar\omega_0$ and $\hbar\omega$ are the excitation and emission photon energies, respectively; $v_F$ is the Fermi velocity, and $Q_{TF}$ is the Thomas-Fermi (TF) screening wave vector that quantifies the effective carrier screening under the TF approximation[38]. As shown in Figure 3a, the PL spectra from both nanospheres and planar nanosheets can be well reproduced by this model, confirming the validity of the assignment of hot-carrier luminescence. Coincidentally, we notice that the value $v_F Q_{TF} \approx \omega_0$ for the excitation flux of 1.47 mJ/cm$^2$. In this case, Equation (1) can be approximated as $I(\omega) \sim \frac{1}{\omega} + \frac{2(\omega_0 - \omega)}{\omega^2}$, where $I(\lambda) \sim \lambda$ in the emission wavelength ($\lambda$) range close to the incident wavelength, leading to a near-linear wavelength dispersion in the hot-carrier luminescence spectra (Figure 3a).

Compared with the nanosheets, the emission intensity from the nanospheres is over 10 times larger (Figure 3a). Moreover, hot-carrier luminescence from nanospheres shows an anomalous dependence on the excitation power, which is also different from the luminescence from GO samples without significant power-dependent shift of emission spectra. The emission intensity at a longer wavelength shows a higher-order dependence on the excitation fluence (Figure 3b) (i.e., a redshift with increasing incident power). This is in sharp contrast to the blueshift measured in graphene nanosheets (Figure 3c) or in the previous reports on exfoliated graphene layers[33]. Such divergence may be resulted from different hot-carrier dynamics in graphene nanospheres. For a better understanding, we compare the power-dependent data from nanospheres and planar nanosheets with the established theoretical model (Eq.1). The fitted parameter of TF screening wave vector $Q_{TF}$ is plotted as a function of the excitation fluence in Figure 3d. Similar to the result for the exfoliated graphene layers[33], the value of $Q_{TF}$ in nanosheets increases with the excitation power monotonically. However, the power-dependence of $Q_{TF}$ for nanospheres is different and it drops with an increased excitation power (Figure 3d).

In theory, the carrier screening effect has been intensively investigated in the past few years. The screening vector $Q_{TF}$ in monolayer graphene is predicted to increase with carrier density ($n$)[38]. The case for multilayer graphene is more complicated due to the interlayer hopping[41-43]. The decrease of $Q_{TF}$ under high-power excitation observed here (Figure 2d) is in consistent with a model of two-dimensional electron gas for multilayer graphene[41]. The effective screening vector $Q_{TF}$ describes the magnitude of carrier screening. The decrease of $Q_{TF}$ in rGO nanospheres indicates an enhancement of carrier scattering, which should be associated to the enhancement of hot-carrier luminescence in nanospheres.

Quantitatively, in monolayer graphene, the carrier-density dependence of $Q_{TF}$ can be expressed as $Q_{TF} \propto n^{1/2}$ [38], which can well explain the divergence of $Q_{TF}$ from the value reported by Liu et al.[33]. For instance, the value of $Q_{TF}$ is approximated to be 0.4 A$^{-1}$ here with the excitation flux of 1 mJ/cm$^2$ in the monolayer nanosheets while the value in Liu's work was reported to be 0.065 A$^{-1}$ with the excitation flux of 1 μJ/cm$^2$ [33]. The carrier density can be estimated with $n \propto \alpha I_{EX} = \alpha_0 I_{EX} / (1 + I_{EX} / I_S)$ with the excitation flux $I_{EX}$, the saturation flux level $I_s$ and the linear absorption coefficient $\alpha_0$. Considering the saturation absorption effect with the reported value of $I_s$ [49], the photo-excited carrier density in our study (1 mJ/cm$^2$) is about 52 times of that in the reference (1 μJ/cm$^2$) [33]. Such a difference corresponds to the ratio of $Q_{TF}$ to be ~ 7.2 in theory, which is close to the experimental value of 6.15 between our work and the literature. As systematically studied by Min et al.[41, 50], our samples of multilayer rGO nanospheres with random stacking order correspond to the case with a chirality of $J \sim 5$. The photo-excited carrier density here is in the order of $10^{13}$ cm$^{-2}$ for the rGO nanospheres, and the fitted value of $Q_{TF} a$ ($a$ is the lattice constant) is in the order of $10^0$ which is in a good agreement with the theoretical prediction[41]. The above results suggest that the screened Coulomb interaction picture provides a good description for the hot-carrier luminescence in rGO nanospheres.

**Photoluminescence dynamics.** To gain more insight into the intrinsic mechanism for the hot-carrier luminescence, we further probe the emission dynamics with a two-pulse correlation measurement. The emission intensity is recorded as a function of the temporal separation between the two pulses. For clarity, we plot the differential intensity (i.e., the intensity difference between the emission at certain representative temporal separation and that at a very long time separation (~ 10 ps)) in Figure 4a. The full-width-at-the-half-maximum (FWHM) of the response curve is ~ 30 fs broader than the pulse correlation curve measured with the second-harmonic response of a BBO crystal (Figure 4a, inset), indicating that the emission arises from an electron-hole distribution within ~ 30 fs post excitation[33]. The FWHM of the correlation curve for nanospheres is relatively smaller in comparison to that for nanosheets (Figure 4a, inset), which is probably a signature of faster build-up of the nonequilibrium population. To quantify the emission dynamics, we analyze the decay curves with a multi-exponential function. The response curves are well-fitted by a bi-exponential decay function with a dominant fast process (~ 0.16 ps) and a slower component (~ 0.77 ps). Previously, an extremely fast process has been assigned to the *e-e* interaction[16, 35]. Typically, the *e-e* interaction occurs in the 10-fs time scale, which cannot be resolved in the current experiment since the pulse duration is broader than such time scale. However, the interaction between electrons and strongly-coupled optical phonons (SCOPs) is in the timescale of < 0.2 ps[16]. Thus, the observed fast component (~ 0.16 ps) is likely to have contributions from both the *e-e* interactions and the electron-SCOP interactions. The relatively slower component (~ 0.77 ps) can be attributed to further equilibration of hot carriers with other phonons.

In Figure 4b, the differential intensity is plotted as a function of wavelength and temporal separation between two pulses. The spectral dispersion of differential intensity at zero delay is similar to the case of steady PL (Figure 4c). We have carefully analyzed the decay lifetime, but no significant wavelength dependence is observed (Figure 4d). In the steady PL, the emission intensities at longer wavelengths grow more rapidly with increasing excitation power, so we check the power-dependent emission dynamics at longer wavelengths (600 - 750 nm). Figures 4e and 4f plot the two-pulse correlation results under relatively low and high power excitation, respectively. Despite the near equal relaxation lifetimes for the two components, the amplitude of the slow component increases significantly under high-power excitation. Since the slow component is connected to the interaction between electrons and weakly-coupled phonons, this result suggests that more electronic energy dissipates with the activation of phonons at higher carrier density.

**Discussion**

The different behaviors between hot-carrier luminescence in spherical nanospheres and planar nanosheets should be related to the structure difference. As discussed in previous literatures[37,51,52], slight bending of graphene plane can dramatically modify the electron dynamics with multiple effects including the decrease of distance between atoms, the rotation of orbitals, as well as rehybrization between the orbitals. These effects can cause additional carrier scatterings when ripples appear in the graphene layers[52]. In rGO nanospheres, such scattering enhancement should be significant since the plane is curled. The enhanced scattering is favorable to promote the hot-carrier luminescence by effectively establishing nonequilibrium carrier population.

We now focus our attention on the power-dependent emission properties in the nanospheres and try to understand the spectral redshift and enhanced *e-ph* interactions with increasing excitation power in the steady and dynamic optical experiments. The increase of carrier scattering (i.e., the reduction of $Q_{TF}$ (Figure 3d) ) with carrier density is not directly related to the spectral redshift (Figures 3b & 3c). Here, we consider the impact of geometrical feature on the carrier scattering in nanospheres. A comprehensive survey on this issue is very challenging and beyond the scope of this work. For a qualitative understanding, we can take a look at the simplest case with two-electron interaction. As illustrated in Figure 4g, the in-plane scattering between two electrons in a nanosphere always companies with a displacement of atom to ensure the conservation of momentum, leading to the extra activation of phonons. This process becomes more pronounced with the enhancement of carrier scattering when the carrier density increases. With more energy loses to the phonon modes, hot-carrier luminescence exhibits a spectral redshift. This is also in consistent with the result of emission dynamics that the slow component increases with carrier density (Figures 4e & 4f). Moreover, the process of supercollision may also increase the magnitude of the *e-ph* interaction. As reported very recently[45, 46], the supercollision cooling of hot carriers induced by three-body interactions between carriers and both phonons and impurities has been found to be important in graphene. Such disorder-assisted scattering process[39] may be more significant in nanospheres since curvature can introduce disorders to an ideal 2D plane. The increased electron temperature with increasing excitation power is favorable for the supercollision cooling, which can strengthen the *e-ph* interactions. Despite many differences between the above discussed two processes, the phonons involved are low frequency ones (e.g., the acoustic phonons[45, 46, 53, 54]), so it is not surprising that no extra peaks appear in the Raman spectrum (Figure 1c).

The above discussions and the model described by Eq. 1 are mainly based on the assumption that the up-converted emission is produced by scattering of carriers excited via linear optical absorption (i.e., inset of Figure 3a)[16, 33, 35]. Soon after the model was developed, efficient nonlinear optical absorption was observed in graphene-based materials.[55, 56] In principle, the nonlinear optical processes, like two-photon absorption (TPA), can also generate carriers that emit light in the up-converted frequency regime as schematically shown in Figure 5a[30]. We have conducted a pump-probe measurement with two different probe wavelengths to indentify the major excitation process. With the pumping wavelength at 800 nm, the carrier dynamics is detected by two probes with the central wavelengths at 700 nm and 600 nm, respectively (Figure 5b). The photo-induced bleaching signal exhibits a bi-exponential decay behavior (Figure 5c). By determining the time delay between the peaks probed at the two wavelengths, we can analyze the population build-up as introduced by Ruzicka *et al*.[57] The differential transmission peak probed at 600 nm is 20-40 fs later than that probed at 700 nm (Figure 5c, inset). For TPA excitation at 800 nm (Figure 5a), the electron-hole pairs are generated with the energy of ~ 3.1 eV (400 nm) which will relax from the high energy level to the lower one, and the peak probed at 600 nm should arrive earlier. For linear optical absorption (inset, Figure 3a), the carrier scatterings dominate the build-up of population at the up-converted frequency range, so that the peak probed at 600 nm should be later than that at 700 nm. The pump-probe data provide unambiguous evidences that carrier scattering is the major excitation channel.

In summary, we have observed the promotion of hot-carrier luminescence in rGO nanospheres in the up-converted frequency regime with fs pulse excitation, which can be well explained in the scheme of screened Coulomb interactions. In the spherical rGO nanostructures, the carrier scattering is significantly enhanced with reduced carrier screening at high carrier density, promoting the establishment of nonequilibrium carrier population. In contrary to the blueshift observed in planar sheets, an abnormal redshift of emission spectrum with increasing carrier density is observed in the nanospheres. This unique property is associated with the strengthened *e-ph* interaction as indicated in the emission dynamics, which may be induced by the processes of extra phonon activation and supercollision cooling in the nanospheres. The efficient up-converted emission in rGO nanospheres can be potentially applied in optical labeling and imaging. This work suggests a strategy to manipulate the interaction of hot carriers by material design for developing efficient graphene-based light sources.

**Methods**

**Sample Preparation and Characterization.** The average diameter of PMMA spheres used here is 300 nm. The cationic polyelectrolyte (PEI) is purchased from Alfa Aesar. GO nanosheets are prepared using a modified Hummer's method from graphite powders. The obtained GO nanosheets are dispersed in deionized water by ultrasonic treatment prior to being used in the experiment. The reference sample of graphene nanosheets is purchased from a commercial source (ACS Material) with purity of ~99.8% and a single-layer ratio of ~ 80%.

To prepare rGO nanospheres, we follow a previous procedure[4]. We modify the surface of PMMA spherical templates with disperser of PEI in aqueous solution. PMMA spheres (5 mg/mL) are dispersed in deionized water containing PEI (2.5 mg/mL) at pH 9.0 under stirring to introduce positive charges at the surface of PMMA spheres. The pH value in the solution is adjusted by dilute HCl or ammonia. After 10 minutes ultrasonic treatment, the suspension is stirred for another 15 minutes to ensure saturated absorption of PEI. To remove the excess PEI, the suspension is further treated by two cycles of centrifugation (8000 rpm at 17% for 5 min) and washing. The PEI-coated PMMA spheres are then dispersed in deionized water and ultrasonically treated for another 10 min. We add a portion of the colloidal suspension of negatively charged GO nanosheets to the turbid PMMA suspension and keep stirring until the supernatant is almost transparent. The sediments are the products owing to the electrostatic interaction of the oppositely charged nanosheets and PMMA surface. The resulting materials are recovered by a separation and washing process, and then dispersed in water by sonication and dried by lyophilization. We place 0.1 g of dried sample in a carbon-powder-surrounding crucible inside a conventional microwave oven in an argon (Ar) atmosphere. The microwave oven (Galanz G80F20CN1L-DG (SO)) functions at full power (800 W, 2.45 GHz) in 200 s cycles (on for 150 s, off for 50 s) for a total reaction time of 600 s. The microwave irradiation reduces GO to graphene, removes the PEI moiety, and decomposes PMMA particles. After reaction, hollow rGO nanospheres are obtained, and trifle PMMA residue is removed with tetrahydrofuran.

The morphology of nanospheres is characterized with TEM (Tecnai F20 S-TWIN, FEI). Crystal structure is characterized with Raman spectroscopy (HR-800, Jabin-Yvon).

**Steady and Transient Optical Measurement.** For optical measurements, same amount of the sample of nanospheres and the reference sample of nanosheets are dispersed in separate PMMA solutions. The films are then prepared by spin coating on silica substrates. Under our experimental configuration, the emission from PMMA matrix is negligible in the wavelength range that we are interested in. The pump laser source is a Ti:sapphire regenerative amplifier (Libra, Coherent Inc.). The laser emits pulses with temporal duration of ~ 90 fs at a repetition rate of 1 kHz. For steady-state PL characterization, the pulses are focused onto a spot with diameter of ~ 200 um using an objective lens (10x, Mitutoyo). The emission light is collected by the objective lens and routed to a spectrometer (Spectrograph 2500, Princeton Instrument) equipped with a liquid-nitrogen-cooled charge-coupled-device (CCD, Spec-10, Princeton Instrument). An ultrasteep optical short-pass edge filter (SP01-785RS, Semrock) is used to exclude the excitation residual. For the two-pulse correlation measurements, two collinear excitation beams are set to be perpendicularly polarized to avoid coherent artifacts. The correlation data are measured with either a photomultiplier tube for traces or the spectrometer for spectra as a function of temporal separation between the two pulses set by a linear translation stage. For the pump-probe studies, we pump the sample at 800 nm and probe the sample with 600 nm or 700 nm generated by an optical parametric amplifier (Opera Sola, Coherent). The time delay between pulses at 600 nm and 700 nm has been precisely calibrated with errors less than 5 fs. The signal is analyzed with a lock-in amplifier and acquired with a Labview program.

**Data Analysis.** The wavelength dispersions of the diffraction efficiency of gratings and the CCD responsivity are calibrated for all spectral measurement. The steady-state PL is analyzed with the theoretical model by fitting the curves to Eq.1., where we take the value of Fermi velocity as $10^6$ m/s approximately. The normalized dynamic curves acquired by either two-pulse correlation are analyzed with a biexponential function in the form of $f(t) = A_1 e^{-t/\tau_1} + A_2 e^{-t/\tau_2}$.

**Acknowledgements**

This work is supported by the National Basic Research Program of China (2012CB921801 and 2013CB932903, MOST), the National Science Foundation of China (91233103, 61108001, 11227406 and 11021403), and the Program of International S&T Cooperation (2011DFA01400, MOST). The author C.Z. acknowledges financial support from New Century Excellent Talents program (NCET-09-0467), Fundamental Research Funds for the Central Universities, and the Priority Academic Program Development of Jiangsu Higher Education Institutions (PAPD).

**Author Contributions**

C.Z. and M.X. conceived the experiments; Q.C., F.X., and W.L. performed the experiments; Y.Z., W.T., and Z.Z. were involved in the sample preparation; Q.C., C.Z., F.X., Y.W. and X.W. analyzed the data; C.Z., and M.X. co-wrote the manuscript with help of all other authors.

**Additional Information**

Competing financial interests: The authors declare no competing financial interests.

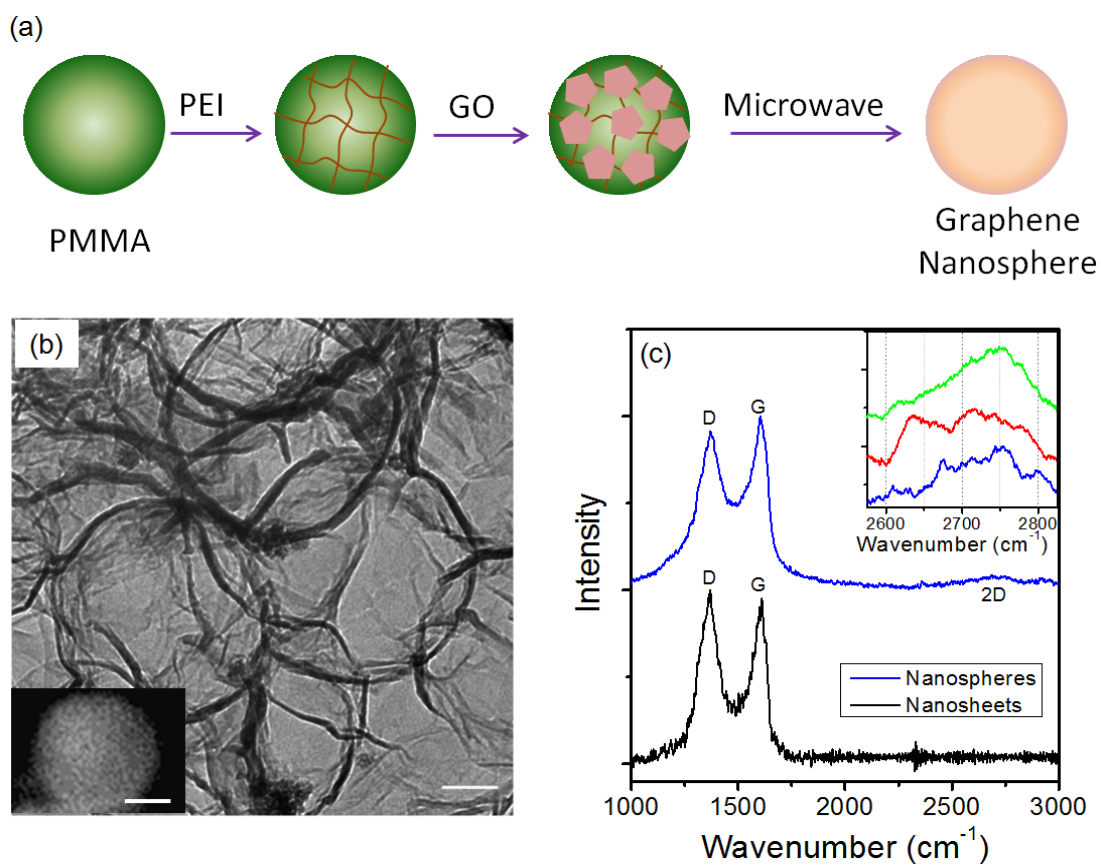

Figure 1, **Preparation and characterization of the graphene nanospheres.** (a) A schematic diagram of the synthesis procedure for the sample of graphene nanospheres. (b) Typical TEM image of the rGO nanospheres (scale bar: 100 nm). Inset shows a SEM image of a rGO nanosphere (scale bar: 100 nm). (c) Raman spectra of the graphene nanospheres and the reference sample of nanosheets. Inset the 2D-mode recorded at the different sample spot.

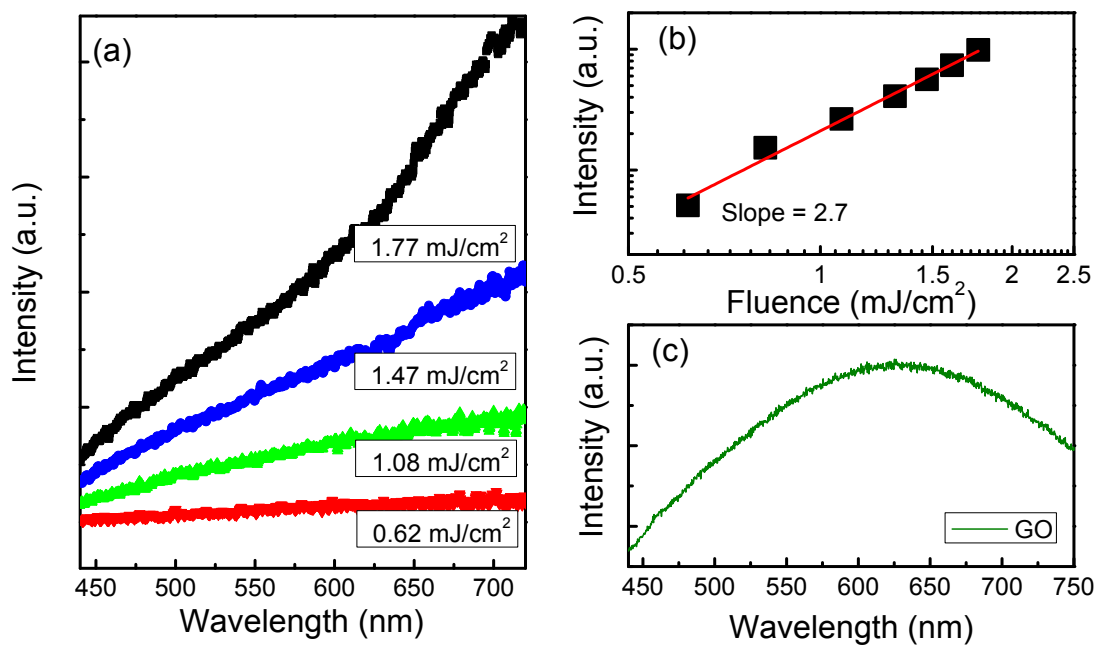

Figure 2, **Hot-carrier luminescence from the graphene nanospheres.** (a) Emission spectra recorded from the graphene nanospheres with fs pulse excitation at 800 nm. (b) The integrated intensity of up-converted emission as a function of excitation fluence. (c) Emission spectrum from graphene oxide excited at 400 nm.

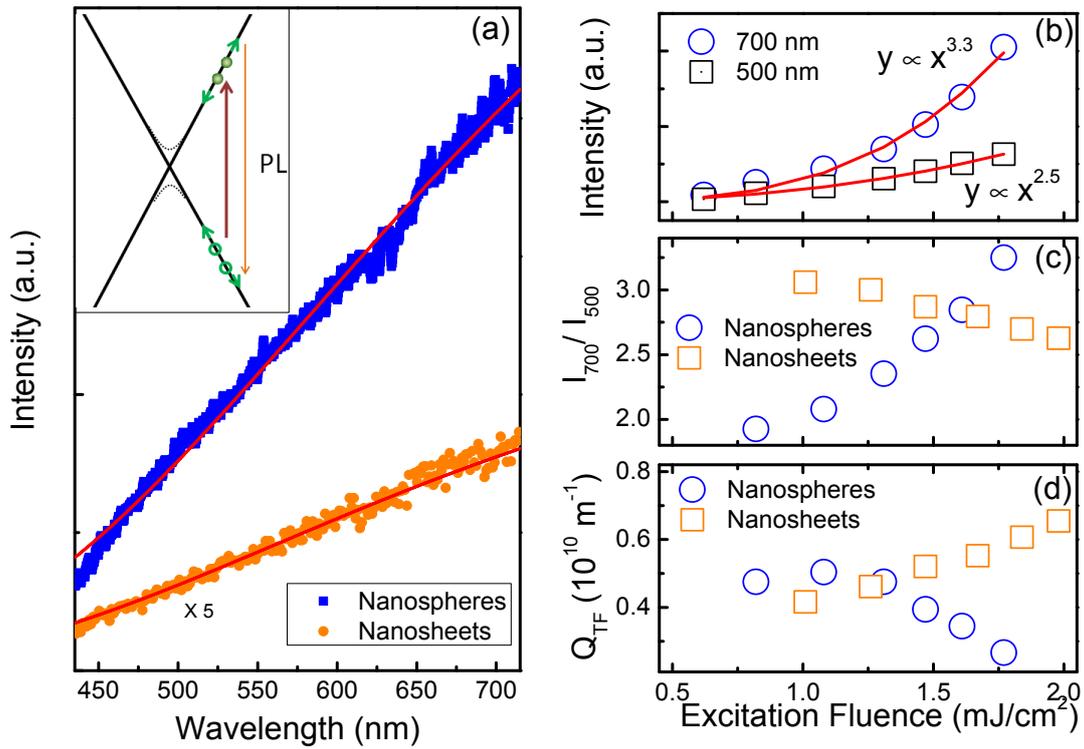

Figure 3, **Comparison the luminescence properties of the graphene nanospheres and nanosheets.** (a) Emission spectra from the samples of graphene nanospheres and nanosheets, respectively, with the excitation fluence of 1.47 mJ/cm$^2$. The solid lines are the curves fitted to the theoretical model (Eq.1). Inset is a schematic diagram of the scattering-induced hot luminescence emission. The dashed line indicates possible modification of band structure. (b) The emission intensities at 500 nm and 700 nm are plotted as a function of the excitation fluence. The solid lines are the curves fitted to different power functions. (c) The ratios between emission intensities at 700 nm and 500 nm, recorded from nanospheres and nanosheets, are plotted as a function of the excitation fluence, respectively. With increasing the excitation pulse, a blueshift for the emission from nanosheets is observed in contrast to a redshift for the emission in nanospheres. (d) Fitted values of the effective screening vector $Q_{TF}$ vs excitation fluence for measurements on the nanospheres and nanosheets, respectively.

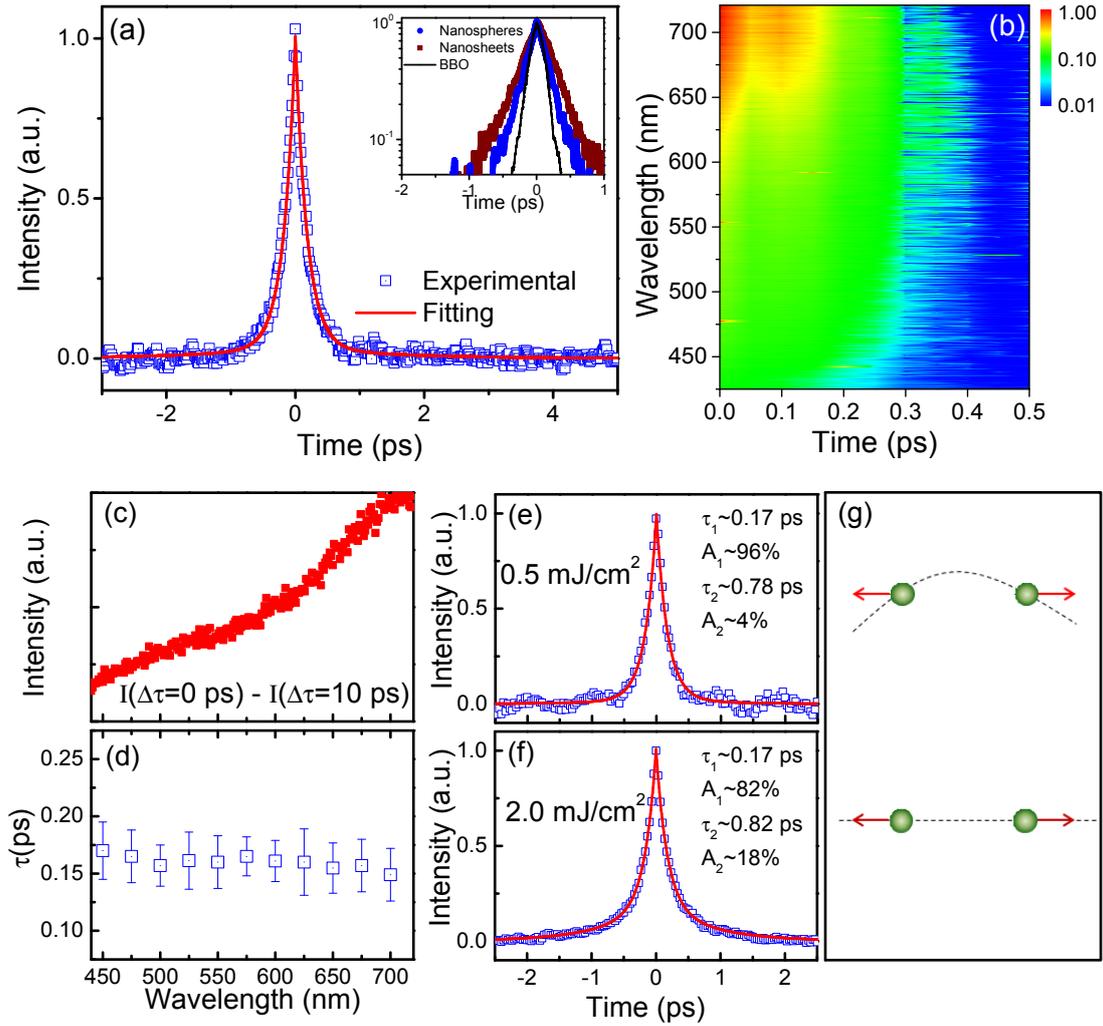

Figure 4, **Two-pulse correlation study on the hot-carrier luminescence of the graphene nanospheres.** (a) The dependence of radiant intensity (450 - 750 nm) on the temporal separation between two excitation pulses (the excitation fluence of each beam is ~ 1.08 mJ/cm$^2$). The intensity at long temporal decay (10 ps) is subtracted for clarity. The solid line is fitting curve to the biexponential function ($\tau_1$ ~ 0.16 ps, $A_1$ ~ 95%, $\tau_2$~ 0.77 ps, $A_2$ ~ 5%). Inset compares the correlation curves from the nanospheres and the nanosheets with the pulse autocorrelation in a logarithmic scale. (b) Counter plot of the emission intensity as a function of wavelength and temporal separation between two excitation pulses. (c) The wavelength-dependent differential intensity at zero separation between two pulses. (d) The wavelength-dependent lifetime of the fast component. (e) and (f) are the two-pulse correlation curves of the emission (600-750 nm) from graphene nanospheres under relatively low (0.5 mJ/cm$^2$) and high (2.0 mJ/cm$^2$) excitation powers, respectively. The fitting parameters are ($\tau_1$ ~ 0.17 ps, $A_1$ ~ 96%, $\tau_2$~ 0.78 ps, $A_2$ ~ 4%) and ($\tau_1$ ~ 0.17 ps, $A_1$ ~ 82%, $\tau_2$~ 0.82 ps, $A_2$ ~ 18%). (g) A schematic diagram compares two-body interactions with curved (top) and planar (down) graphene plane.

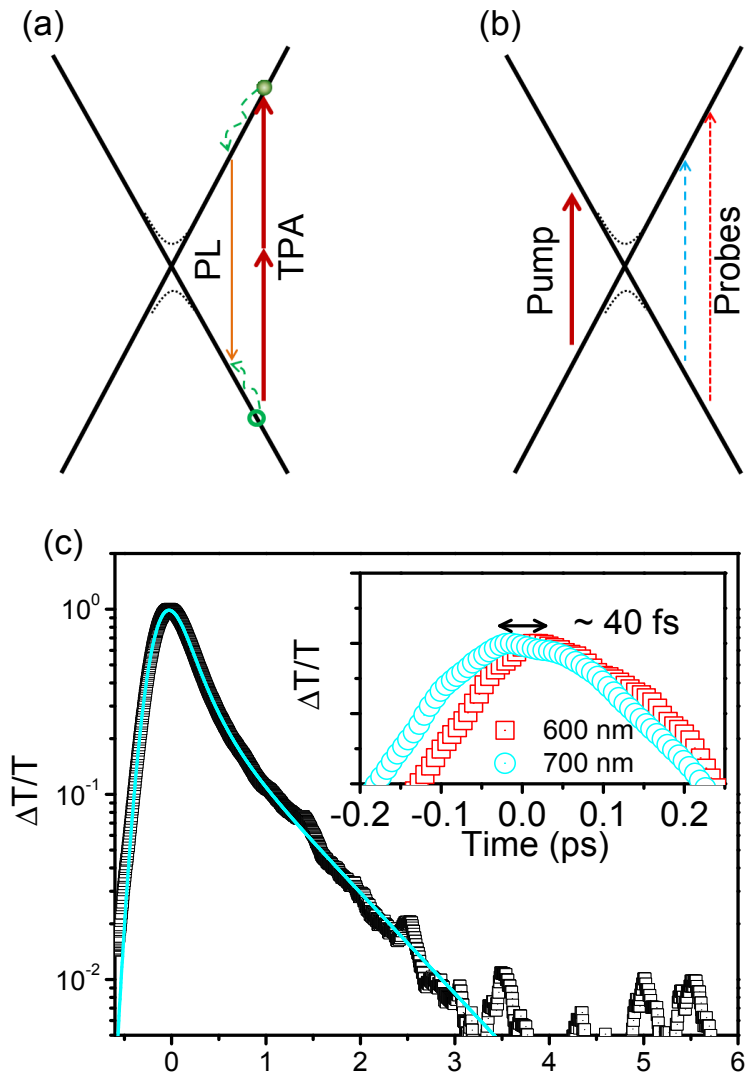

Figure 5, **Pump-probes studies on the possible excitation mechanisms.** (a) The sketch diagram of PL emission induced by the process of TPA. (b) Two-color pump-probe scheme. (c) With pumping at 800 nm, the differential transmission probed at 700 nm is plotted as a function of the decay time. The solid line is a curve fitting to the biexponential function. The inset shows a slight displacement between peaks with probes at 700 nm and 600 nm.